\documentstyle[12pt,fleqn]{article}
\begin{document}
\newtheorem{lemma}{Lemma}
\newtheorem{prop}{Proposition}

\title{On the existence of the second Dirac operator in Riemannian space}

\author{Vladimir V Klishevich}

\maketitle

\begin{center}
{\small Class. Quantum Grav. {\bf 17} (2000) 305-318. Printed in the UK

Department of Physics, Omsk State University, pr. Mira 55a, Omsk, Russia

E-mail: klishe@univer.omsk.su

Web: http://www.univer.omsk.su/ $\widetilde{}$ klishe/index.html}
\end{center}

\begin{abstract}
We describe a Riemannian space class where the second Dirac operator arises
and prove that the operator is always equivalent to a standard Dirac one.
The particle state in this gravitational field is degenerate to some extent
and we introduce an additional value in order to describe a particle state
completely. Some supersymmetry constructions are also discussed.
As an example we study all Riemannian spaces with a five-dimentional motion
group and find all metrics for which the second Dirac operator exists. On the
basis of our discussed examples we hypothesize about the number of second
Dirac operators in Riemannian space.
\end{abstract}


\section{Introduction}

The Dirac equation is one of the most important equations in theoretical
physics. Many articles and monographs have been devoted to the investigation
of this equation and its solutions \cite{Bjorken,Itzykson}.
The generalization of this equation for all Riemannian manifolds is rather
natural and gives us the task of finding the motion of a spin-$\frac{1}{2}$
particle in an external gravitational field. The gravitation considerably
changes the motion dynamics of the particle and new effects arise
that are not observed in flat space. We will discuss one of these effects
that arises because of a second `non-standard' Dirac operator as the literature
on this problems shows (we will give an exact definition later on).
A `non-standard' Dirac operator arises in those Riemannian spaces
in which a symmetric Killing tensor $a_{ij}$ ($a_{(ij;k)}=0$) exist, the
square root of which is defined by an antisymmetric Killing-Yano tensor.
Penrose was the first to draw attention to this fact \cite{Penr}.
We suggest another way to define the second Dirac operator and afterwards
generalize it, reducing the difficulty of evaluating the egenvalue for the
Riemannian space metric. Using our method we find a class of gravitational
fields where the second Dirac operator exists and establish an equvalence
between the second Dirac operator and the `standard' Dirac one.
Further proof of the theorem dealing with the equality of indices of
`standard' and `non-standard' Dirac operators is given (see \cite{nondir}).
We also prove that in spaces where the second Dirac operator exists
there always appears two sequences of differential matrix operators.
Operators of the first sequence commute with the Dirac operator and
operators of the second sequence anticommute with the Dirac operator.
We give explisit equations for this operator and discuss some
supersymmetry constructions. The presence of a second Dirac operator
makes it possible to obtain a second solution of the initial Dirac equation,
so in some sense the spinal particle state appears to be degenerated and
one has to introduce an additional value to describe the state of a particle
completly. We interpret this value as a parity analog.
As an example we give all Riemannian space metrics with a five-dimentional
motion group where the second Dirac operator exist and formulate
a working hypothesis concerning the number of second equivalent
Dirac operators.

In section \ref{s1} we discuss a Dirac operator and a second Dirac
operator in Riemannian space. In this section a Riemannian space class
is discribed where a second Dirac operator arises. In section \ref{s2}
we prove an equivalence between a Dirac operator and a second Dirac one and
study some properties of value of the equivalence between these operators.
In section \ref{s3} we give all space metrics with five-dimentional motion
group where a second Dirac operator exists and in section \ref{s4} we discuss
the solution of the Einstein equation and Petrov types for all discussed
metrics. Our hypotheses concerning the number of second Dirac operators
is given in section \ref{s5}.

\section{When does the second Dirac operator arise? \label{s1}}
\subsection{General remarks on a Dirac operator in Riemannian space}

Let us write a free Dirac equation in some system of coordinates $\{x^i\}$
in {\it four-dimensional} Riemannian space
\begin{equation}
D \Psi \equiv \gamma^k P_k \Psi = m\Psi. \label{dirac}
\end{equation}
Here $D$ denotes a (`standard') Dirac operator, $m$ is a constant,
$P_k = i(\nabla_k + \Gamma_k)$, $\nabla_k$ is a covariant derivative operator,
$\Gamma_k$ is a spin connection, where repeating indices indicates summation.
A spin connection describes the interection of spinors with the gravitational
field, but our choice is not quite unique (see \cite{string}, vol 2).
For the spin connection we use the following equation:
\begin{equation}
[P_k,\gamma_i]=0, \hspace{5mm} {\rm Sp}(\Gamma_i)=0, \label{sviaz}
\end{equation}
If we open the brackets in (\protect\ref{sviaz}) we find the following
equation:
\begin{eqnarray}
\gamma_{k;i}=[\gamma_k,\Gamma_i] \label{spin4}
\end{eqnarray}
here and later $[A,B]=AB - BA$ is a commutator of $A$ and $B$,
$\{A,B\} = AB + BA$ is an anticommutator, the symbol $a_{;k}$ denotes the
covariant derivative with respect to coordinate $x^k$ about a metric,
symbol $a_{,k}$ or $\partial_{x^k}$ denotes a partial derivative with
respect to the coordinate $x^k$.

Dirac matrices in Riemannian space are defined as arbitrary but
fixing a solution of the system
\begin{equation}
\gamma^i\gamma^j + \gamma^j\gamma^i = 2g^{ij}(x){\widehat E}_4, \label{sysg1}
\end{equation}
where ${\widehat E}_4$ is a unit $4 \times 4$ matrix and $g^{ij}$ is a metric
tensor.

Shapovalov proved \cite{Shap1} that the choice of the spin connection
in the form (\protect\ref{sviaz}) yields the existence and uniqueness
(to a precision of type equvalence $S^{-1} \gamma S$, where $S$ is a
non-singular matrix) of the system solution (\protect\ref{sysg1}).
That is why we choose the spin connection in the form (\protect\ref{sviaz}).
If we change $\gamma$-matrices to equivalent matrices
$\stackrel{-}{\gamma} = S^{-1} \gamma S$, the spin connection changes to
$\stackrel{-}{\Gamma}_i = S^{-1} \Gamma_i S + S^{-1}S_{,i}$,
and now the Dirac operator is `covariant' about our transformation
$\stackrel{-}{D} = S^{-1} D S$.

To construct Dirac matrices in Riemannian space in explicit form we use
the tetrade method. For this purpose in $4 \times 4$ matrix space
we introduce a basis
${\widehat E}_4$,
${\widehat\gamma}$,
${\widehat\gamma}^i$,
$\stackrel{*}{\widehat \gamma^i}$,
${\widehat\gamma}^{ij}$.
A matrix ${\widehat\gamma}$ is defined as
${\widehat\gamma}= - {\widehat\gamma}^1{\widehat\gamma}^2{\widehat\gamma}^3{\widehat\gamma}^4$,
matrices $\stackrel{*}{\widehat \gamma^i} = - \frac{1}{3!} \varepsilon_{ijkl}{\widehat\gamma}^j {\widehat\gamma}^k {\widehat\gamma}^l$
and ${\widehat\gamma}^{ij}=\frac{1}{2}[{\widehat\gamma}^i,{\widehat\gamma}^j]$.
Let us define $\gamma^i(x)=e_{(a)}^i{\widehat \gamma}^a$,
where ${\widehat \gamma}^a$ are constant Dirac matrices (with a hat).
An explicit form for these matrices can be found in appendix A.
$e_{(a)}^i$ is the orthogonal tetrad, these values split the constant
metric $e^i_{(a)} e_{(b)i}=\eta_{ab}$, here $\eta_{ab}$ is the Minkovsky
matrix, which is a constant diagonal matrix. Indices are raised and lowered
by $\eta$ for tetrad indices, and by $g$ for coordinate indices. All
indices run from 1 to 4. We choose a signature of the form
$\eta_{ab}=\eta^{ab}={\rm diag}(+1,-1,-1,-1)$. All tetrades for our examples
can be found in appendix B.

For the spin connection one can use the following equations:
\begin{equation}
\Gamma_i = - \frac{1}{4} \gamma_{k;i}\gamma^k \hspace{5mm} {\rm or} \hspace{5mm} \Gamma_i = \frac{1}{4} \gamma^k\gamma_{k;i}. \label{sviaz2}
\end{equation}
These equations are derived on the base of the following:
\begin{equation}
P_k\gamma_i = 0. \label{sviaz3}
\end{equation}
If we open the brackets in (\protect\ref{sviaz3}) we find (\protect\ref{sviaz2}).
Spin connections of the form (\protect\ref{sviaz2}) are often called
Fock-Ivanenko coefficients. We use equations (\protect\ref{sviaz2}) for our
examples.
\begin{lemma}
{\it Equations (\protect\ref{sviaz}) and (\protect\ref{sviaz3}) in four-dimensional
Riemannian space are equivalent.}
\end{lemma}
Proof. We prove that equation (\protect\ref{sviaz3}) follows from
equation (\protect\ref{sviaz}), we seek the form
\begin{displaymath}
\Gamma_i = a_i \gamma + b^n_i \gamma_n + c^{nm}_i \gamma_{nm} + d^n_i \stackrel{*}{\gamma}_n.
\end{displaymath}
We substitute this expansion into (\protect\ref{sviaz}) and equate to zero
the coefficients on the bases $\widehat\gamma$-matrices and obtain the
following equation (with $c^{nm}_i=-c^{mn}_i$)
\begin{displaymath}
\gamma_{k;i} = 4 c^{nm}_i g_{nk} \gamma_m, ~~~ a_i=b^n_i=d^n_i=0.
\end{displaymath}
Multiplying this equation from the left by $\gamma^k$ and contracting the
index $k$ we find
\begin{displaymath}
\gamma^k\gamma_{k;i} = 4 c^{nm}_i \gamma_n \gamma_m = 4 \Gamma_i.
\end{displaymath}
Here we use the equation $\gamma_i\gamma_j=\gamma_{ij}+g_{ij}{\widehat E}_4$,
so
\begin{equation}
\Gamma_i = \frac{1}{4}\gamma^k\gamma_{k;i}. \label{spin5}
\end{equation}
Comparing equations (\protect\ref{sviaz2}) and (\protect\ref{spin5}), we
deduce that in four-dimensional space these connections are equal.
The second condition in (\protect\ref{sviaz}) is obvious.

{\it Remark}. Conditions (\protect\ref{sviaz}) and (\protect\ref{sviaz3})
are not equvalent in other dimensions.

A general form of the symmetry operators for the equation (\protect\ref{dirac})
was found by Shapovalov \cite{Shap1}. This problem was also solved
independently in papers by Carter and McLenaghan \cite{CarMac} and
McLenaghan and Spindel \cite{MacSpi,MacSpi2}. The symmetry operators
for Dirac equation (\protect\ref{dirac}) has the following form:
\begin{eqnarray}
L & = & {\widehat E}_4 \xi^k P_k - \frac{i}{4} \gamma^{kl} \xi_{k;l} \label{sym1}\\
L & = & \stackrel{*}{\gamma_j}f^{kj}P_k + \frac{i}{3}\gamma_j\widetilde f^{kj}{}_{\!;k}, \label{sym2}\\
L & = & 2\gamma \gamma^{lk}f_kP_l + \frac{3i}{4} \gamma f^k{}_{\!;k}. \label{sym3}
\end{eqnarray}
Here
$\gamma ^{kl}=\frac{1}{2}[\gamma^k,\gamma^l]$,
$\gamma = - \frac{1}{4!} e_{ijkl}\gamma^i \gamma^j \gamma^k \gamma^l$,
$\stackrel{*}{\gamma_i} = - \frac{1}{3!}e_{ijkl}\gamma^j \gamma^k \gamma^l$,
$\widetilde f_{ij} = \frac{1}{2} e_{ijkl}f^{kl}$,
$e_{ijkl} = \sqrt{ - \det(g_{mn})}\varepsilon _{ijkl}$, a complete
antisymmetric tensor $(\varepsilon _{1234}=1)$.

The vector field $\xi^k$ in operator (\protect\ref{sym1}) is defined from the
equations $\xi_{i;k} + \xi_{k;i} = 0$ and is called the Killing vector field.
A tensor field in operator (\protect\ref{sym2}) is found from equations
\begin{equation}
f_{ij} + f_{ji} = 0, ~~ f_{ij;k}=e_{ijkl}\,g^l, ~~ (f_{ij;k}+f_{ik;j}=0), \label{kili}
\end{equation}
where $g^l$ is a vector and is called the Killing-Yano tensor.
A vector field in operator (\protect\ref{sym3}) is defined by the equation
$f_{i;j} = \frac{1}{4} g_{ij}f^k_{~;k}$, $f_i=\eta_{,i}$ where $\eta$ is a scalar
and is called the Yano vector field.
Equations for this field are found, for instance, in \cite{Shap1,CarMac,MacSpi}.
Operators (\protect\ref{sym1}), (\protect\ref{sym2}) and (\protect\ref{sym3})
commute with a Dirac operator (\protect\ref{dirac})
according to the definition $[D,L] = 0$.

\subsection{Second Dirac operator}

Carter and McLenaghan \cite{CarMac} and McLenaghan and Spindel
\cite{MacSpi,MacSpi2}
showed that using the Killing-Yano tensor field we can construct
a symmetric Killing tensor:
\begin{equation}
a^{ij} = f^{ik}f^j_{~k}. \label{kiling1}
\end{equation}
This tensor together with Killing vectors under some additional conditions
makes a complete commutative set for the Klein-Fock equation \cite{Bagrov}.
We define a Klein-Fock operator as a quadrate of the Dirac operator (the
equality under derivatives of higher order)
\begin{equation}
K = D^2. \label{KF}
\end{equation}
It is remarkable that there are Riemannian spaces where for a symmetry
operator (\protect\ref{sym2}) the following condition holds:
\begin{equation}
L^2 = D^2, \hspace{5mm} L \ne D  \label{kiling4}
\end{equation}
and our aim is to describe this class of gravitational fields.
We interpret the symmetry operator $L$ (\protect\ref{sym2}) under condition
(\protect\ref{kiling4}) as the second (`nonstandard')
Dirac operator.
The equality (\protect\ref{kiling4}) we view as an identity (in a weaker
sense than in equation (\protect\ref{KF}), but we shall prove that equation
(\protect\ref{kiling4}) always holds identically).
Let us write out values under derivatives in equations (\protect\ref{dirac})
and (\protect\ref{sym2}) and quadrate. Using the tetrad technique we see that
if (\protect\ref{kiling4}) is fulfiled the following conditions are to be
satisfied:
\begin{equation}
 f^{ik}f^j_{~k} = g^{ij}. \label{kiling3}
\end{equation}
that is, Killing tensor (\protect\ref{kiling1}) should be equal to
metric tensor.

One can see equation (\protect\ref{kiling3}) as an egenvalue problem:
\begin{equation}
 A_{injm} g^{nm} = \lambda g_{ij}, \hspace{5mm} A_{injm} = f_{in}f_{jm}, \label{form}
\end{equation}
since a Killing-Yano tensor field is defined to the precision of any
multiplier, the parameter $\lambda$ has, in principle, only two meanings:
$\lambda = 0,1$. In the case where $\lambda = 1$ the second Dirac operator
arises. It is necessary to have an individual investigation in the case
where $\lambda = 0$ because in this case there is a symmetry operator
(\protect\ref{sym2}) under the condition $L^2 = 0$. We shall give some
examples of this gravitational field later on.

A basic property of the Killing-Yano tensor field in equation
(\protect\ref{form}) is given as the following proposition.
\begin{prop}
{\it A Killing-Yano tensor field satisfying condition (\protect\ref{form})
is covariantly constant.}
\end{prop}
Proof. {\it Case $\lambda=1$}. First of all it is necessary to point out that
from equation (\protect\ref{form}) considered as multiplication of matrices
it follows that $\det(f_{ij}) = \pm \det(g_{ij}) \ne 0$. Contracting
equation (\protect\ref{kiling3}) over indices $i$ and $j$ and taking a
covariant derivative we have that $2f^{ik}f_{ik;n} = 0$. Using equation
(\protect\ref{kili}) we then have ${\widetilde f}_{nl}g^l = 0$. Wiewing this
equation as a linear homogeneous system of equations for unknown functions
$g^l$ under the condition $\det({\widetilde f}_{nl}) \ne 0$ we deduce that
all $g^l = 0$ and (see equation (\protect\ref{kili}) again) we see that
$f_{ij;k} = 0$.

{\it Case $\lambda=0$}. Let us contract the equation
$0 = f^{ik}f_{ik;n} = f^{ik}e_{iknl}g^l$ with the tensor $e^{stnl}$ by
index $n$ we then have that $g_rf_{st} + g_sf_{tr} + g_tf_{rs} = 0$.
We contract this equation ($g_rf_{st} + g_sf_{tr} + g_tf_{rs}$)
with the tensor $e^{stnl}$ again. Thus $g_r{\widetilde f}_{nl}=0$
that is all $g_r=0$ and $f_{ij;k}=0$ as well.

{\it Remark.} The proof in the case where $\lambda=0$ is also appropriate
in the case where $\lambda=1$.

So a Riemannian space class where the second Dirac operator arises
is defined by the following conditions:
\begin{equation}
g^{ij} = f^{ik}f^j_{~k}, \hspace{5mm} f_{ij;k} = 0, \hspace{5mm} f_{ij} + f_{ji} = 0. \label{kiling3'}
\end{equation}
For more detailed information one can use a space classification admitting
a covariant constant tensor structure and (see, for example, \cite{nondir2}).
A systematic classification of manifolds admitting solutions
(\protect\ref{kili}) is seen in \cite{rudi1,rudi2} and
partially in \cite{Shap1}. One can use these papers but the conditions
(\protect\ref{kiling3'}) are formulated in covariant form and are very
suitable for practice calculations.

{\it Remark.} If $f_{ij;k} = 0$, then ${\widetilde f}_{ij;k} = 0$ of course,
so that the `tail' in operator (\protect\ref{sym2}) is vanishes and the
equality (\protect\ref{kiling4}) becomes an identity.

{\it Remark.} If $f_{ij;k} = 0$, then some coordinate system exists
where the matrix $||f^i_{~j}||$ is constant
(is reduced to normal Jordan form) \cite{ShirokAP}.
In this case we can write equation (\protect\ref{form}) in the following form:
\begin{equation}
 A^{nm}_{ij} g_{nm} = \lambda g_{ij}, \hspace{5mm} A^{nm}_{ij} = f^n_{~i}f^m_{~j}
\end{equation}
and without the restriction of generality we assume the matrix $A^{nm}_{ij}$
to be constant.

\section{What does the existence of a second Dirac operator give? \label{s2}}

\begin{lemma}
{\it If for Dirac operator (\protect\ref{dirac}) the condition
(\protect\ref{kiling4}) is carried out then a nonsingular matrix $S$ exist
and}
\begin{equation}
 SL = DS. \label{equiv}
\end{equation}
\end{lemma}
Proof. The choice of the spin connection in the form (\protect\ref{sviaz})
yields the uniqueness (to a precision of type equvalence $S^{-1} \gamma S$,
where $S$ is a non-singular matrix) of the system solution
(\protect\ref{sysg1}) (Shapovalov's result \cite{Shap1}).

So that in seeking a solution of the Dirac equation we can use the operator
$L$. Moreover, this equation answers the question about the index of the
operator $L$
\begin{displaymath}
{\rm index}(L) = {\rm index}(D),
\end{displaymath}
because $\det S \ne 0$ ($L$ is a Dirac operator in the other representation).
So that the term `nonstandard' in our opinion can be referred not to the
operator $L$ but probably to the gravitational field itself.
I would like to draw your attention to equation (\protect\ref{kiling3}).
An antisymmetric tensor (\protect\ref{kili}) splits a metric anologouisly
to the metric itself ($g^{ik}g^j_{~k}=g^{ij}$).

The existence of the second Dirac operator gives some additional interesting
information. First, we note that from (\protect\ref{kiling4}) and
(\protect\ref{equiv}) it follows that the matrix $S$ is a symmetry for
the quadratic Dirac operator:
\begin{displaymath}
D^2 = DD = (SLS^{-1})(SLS^{-1}) = SL^2S^{-1} = SD^2S^{-1},
\end{displaymath}
and for all even degree of the Dirac operator the matrix $S$ is a symmetry
too: $[D^{2n},S] = 0$. Moreover,
\begin{equation}
[D^2,S] = D[D,S] + [D,S]D = 0 \label{anticom}
\end{equation}
and we see that the operator $D^{(0)}_{+} = [D,S]$ anticommutes with the
Dirac operator $D$. One can see that this operator commutes with the
quadratic Dirac operator (and for all even degree).
As $D^2D^{(0)}_{+} = - DD^{(0)}_{+}D = D^{(0)}_{+} D^2$,
deducing as in (\protect\ref{anticom}) we see that the operator
$D^{(1)}_{+} = [D,D^{(0)}_{+}]$ anticommutes with the Dirac operator $D$.
Let us define by induction an operator $D^{(k)}_{+} = [D,D^{(k-1)}_{+}]$, $k=1,2,3,\ldots$.
All of these operators anticommute with the Dirac operator $D$.
Let us introduce an operator $D^{(0)}_{-} = \{D,S\}$. One can see that
this operator commutes with the Dirac operator. By analogy all operators
$D^{(k)}_{-} = \{D,D^{(k-1)}_{-}\}$, $k=1,2,3,\ldots$ commute with
the Dirac operator as well.

Analogous sequences of operators can be constructed for the operator $L$.
Let us introduce the following operator $L^{(0)}_{+} = [L,S]$, then by
induction $L^{(k)}_{+} = [L,L^{(k-1)}_{+}]$, $k=1,2,3,\ldots$.
All of these operators anticommute with the second Dirac operator $L$
and operators $L^{(0)}_{-} = \{L,S\}$ and
$L^{(k)}_{-} = \{L,L^{(k-1)}_{-}\}$, $k=1,2,3,\ldots$ commute with $L$.

Let us see some properties of the matrix $S$.
\begin{lemma}
{\it Matrix $S$ is unitary.}
\end{lemma}
Proof. As the Dirac operator $D$ (and $L$) is self-adjoint let us assume
Hermitian conjugation of equation (\protect\ref{equiv}) $LS^{+} = S^{+}D$.
Multiplying this equation on the left by $S$ and using (\protect\ref{equiv})
again, we find $[D,SS^{+}] = 0$. Writing out summands from
derivatives in this equation, we deduce that $[\gamma^i,SS^{+}] = 0$,
so for all matrices $A$ equation $[A,SS^{+}] = 0$ is correct. So
the matrix $SS^{+}$ is scalar, i.e. $SS^{+} = \lambda E_4$.
Without the restriction of generality we assume $\lambda = 1$.

Let it be that $D^{(\pm)} = (D \pm L)/2$.
\begin{lemma}
{\it If matrix $S$ is Hermitian ($S^{+}=S$), then $S^2=1$ and $[D^{(+)},S] = 0$ and
$\{D^{(-)},S\} = 0$.}
\end{lemma}
Proof. Let us remark that if $S^2=1$ equation (\protect\ref{equiv})
can be written in the form $LS = SD$.

Equations $L^{(0)}_{+} = - D^{(0)}_{+}$ and $L^{(0)}_{-} = D^{(0)}_{-}$
follow from this lemma, and more general equations
$L^{(k)}_{+} = (-1)^k D^{(k)}_{+}$ and $L^{(k)}_{-} = D^{(k)}_{-}$, because of
equations $D^{(k)}_{\pm} = 2^k D^k D^{(0)}_{\pm}$,
$L^{(k)}_{\pm} = 2^k L^k L^{(0)}_{\pm}$ are correct.

Let us look at the following structure: $\left(D^{(-)},K^{(-)},S\right)$.
This structure has the following properties:
\begin{displaymath}
K^{(-)} = D^{(-)^2}, \hspace{5mm} S^2 = 1, \hspace{5mm} \{D^{(-)},S\} = 0.
\end{displaymath}
In accordance with \cite{Wit} the three members
$\left(D^{(-)},K^{(-)},S\right)$ possess a supersymmetry.
A conjugating structure $\left(D^{(+)},K^{(+)},S\right)$ has the following
properties
\begin{displaymath}
K^{(+)} = D^{(+)^2}, \hspace{5mm} S^2 = 1, \hspace{5mm} [D^{(+)},S] = 0.
\end{displaymath}
So in the Riemannian space class (\protect\ref{kiling3'}) there is a
supersymmetry construction.

Let us prove that the existence of the second Dirac operator gives a state
degeneration of a particle. As the operators $D$ and $L$ commute
let us solve the egenvalue problem:
\begin{displaymath}
D \Psi = m  \Psi, \hspace{5mm} L \Psi = m' \Psi.
\end{displaymath}
However, the condition (\protect\ref{kiling4}) gives $m'= \pm m$ at once and
with (\protect\ref{equiv}) we have
\begin{displaymath}
SL \Psi = DS \Psi = \pm m S\Psi,
\end{displaymath}
so both solutions $\Psi$ and $S\Psi$ may sytisfy
the Dirac equation; in our case $D^{(0)}_{+}=[D,S] \ne 0$ and the solution
$\Psi$ differs from $S\Psi$. Together with the pair of equations
\begin{displaymath}
D \Psi = m  \Psi, \hspace{5mm} DS \Psi = \pm m S\Psi
\end{displaymath}
we can use a conjugate pair
\begin{displaymath}
L \Psi = \pm m  \Psi, \hspace{5mm} LS \Psi = m S\Psi.
\end{displaymath}

Let us look at the two states of the particle $\Psi^{(\pm)}$:
\begin{displaymath}
D \Psi^{(\pm)} = m \Psi^{(\pm)}, \hspace{5mm} L \Psi^{(\pm)} = \pm m \Psi^{(\pm)}.
\end{displaymath}
To find states $\Psi^{(\pm)}$ we introduce the following operators:
\begin{displaymath}
D^{(\pm)} = (D \pm L)/2.
\end{displaymath}
They have the following properties:
\begin{displaymath}
D^{(\pm)} \Psi^{(\pm)} = m \Psi^{(\pm)}, \hspace{5mm} D^{(\pm)} S\Psi^{(\pm)} = \pm m S\Psi^{(\pm)}, \hspace{5mm} D^{(\pm)} \Psi^{(\mp)} = 0, \hspace{5mm} D^{(\pm)} D^{(\mp)} = 0.
\end{displaymath}

If matrix $S$ is Hermitian we deal with some physically observable value.
A condition $S^2=1$ gives us the possibility of interpreting this value as a
parity analogue.

\section{The second Dirac operator in Riemannian space with a five
dimensional motion group \label{s3}}

As an illustration of our theory let us look at the Riemannian spaces with a
five-dimensional motion group given in Petrov's monograph \cite{Petrov}.
Of the $39$ metrics in \cite[p 297-314]{Petrov} only four metrics admit
the second Dirac operator. To find these metrics we solved an equation
for the Killing-Yano tensor fields (\protect\ref{kili}) and checked the
condition (\protect\ref{kiling3}).

Using specific calculations it is rather useful to follow these conditions.
Let us derive equation (\protect\ref{kili}) and exclude the second
and then the first derivatives with the help of equations (\protect\ref{kili}).
Then we shall have an equation in the form
\begin{equation}
(R^n_{~kij} + R^n_{~jik})f_{in} - R^n_{~iik}f_{jn} - R^n_{~iij}f_{kn} = 0, \label{kili_2}
\end{equation}
This equation holds for all indices $i,j,k$.
The Riemann tensor is defined by the equation
$R^i_{~jkl} = \Gamma^i_{jl,k} - \Gamma^i_{jk,l} + \Gamma^i_{nk} \Gamma^n_{jl} - \Gamma^i_{nl} \Gamma^n_{jk}$,
and Ricci tensor by contraction over the first and third indices:
$R_{jk} = R^n_{~jnk}$.
For constant curvature spaces under substitution
\begin{equation}
R_{ijkl} = \frac{R}{n(n-1)}(g_{ik}g_{jl} - g_{il}g_{jk}),~~R_{jl} = \frac{R}{n}g_{jl}~~(n=4) \label{post}
\end{equation}
equation (\protect\ref{kili_2}) is carried out identically, so for spaces
(\protect\ref{post}) these conditions are useless. However, if the space
is not of type (\protect\ref{post}), the conditions (\protect\ref{kili_2})
help the search for equation solutions (\protect\ref{kili}).
In all cases we calculate a matrix $S$ that is equivalent (\protect\ref{equiv}).

Let us look at metric number $(33.31)$ in Petrov's monograph \cite[p 309]{Petrov}.
\begin{eqnarray}
ds^2 & = & 2k_{13} \exp(-cx^4) dx^1 dx^3 + k_{22} \exp(-2cx^4) dx^{2^2} + 2k_{22} x^1 \exp(-2cx^4) dx^2 dx^3 + \nonumber \\
     &   & k_{22} x^{1^2} \exp(-2cx^4) dx^{3^2} + k_{44} dx^{4^2} \label{m33_31}
\end{eqnarray}
where $c$ and $k_{ij}$ are constants, $\det(g_{ij})= - k_{13}^2 k_{22} k_{44} \exp( - 4cx^4)$,
and the space scalar curvature $R = (k_{22}/k_{13}^2 - 11c^2/k_{44})/2$.

Using conditions (\protect\ref{kili_2}) we find a simple connection
between Killing-Yano field components: $f_{14} = f_{12} c \exp(cx^4)k_{12}/k_{22}$.
The case $k_{22}k_{44}+c^2k_{13}^2 \ne 0$, $c \ne 0$ gives the conditions
$f_{12}=f_{14}=f_{23}=f_{24}=0$. Let us now solve equation (\protect\ref{kili})
and deduce that all components are zero $f_{ij}=0$ and the space (\protect\ref{m33_31})
does not admit a Killing-Yano tensor field.
If the condition $k_{22}k_{44}+c^2k_{13}^2=0$ is carried out ($c \ne 0$,
otherwise the metric is degenerate), equations
(\protect\ref{kili_2}) gives an additional condition between the Killing-Yano
field components: $f_{34} = f_{24}x^1 - f_{23} \exp(cx^4)k_{44}/(ck_{13})$.
Using these relationships and solving equations (\protect\ref{kili}) we shall
find that the space (\protect\ref{m33_31}) admits a Killing-Yano tensor field
with non-zero components:
$f_{13} = a\exp(-c x^4)$,
$f_{24} = ca\exp(-c x^4)$,
$f_{34} = ca x^1 \exp(-cx^4)$,
$a=ik_{13}$.
In other cases Riemannian space (\protect\ref{m33_31}) does not admit the
Killing-Yano tensor field. In the case where $c=0$ the space admits
a Yano vector field, but we will not look at this situation.

Let us look at the case where $k_{22}k_{44}+c^2k_{13}^2=0$ ($c \ne 0$).
A tetrad is given in appendix B.

The Dirac operator in the space (\protect\ref{m33_31}) has the form
\begin{eqnarray*}
 D & = & \frac{\exp(cx^4)}{2k_{13}} ({\widehat \gamma}^1 - {\widehat \gamma}^2) \partial_{x^1} + \left(\frac{\sqrt{k_{44}}\exp(c x^4)}{c k_{13}} {\widehat \gamma}^4 - x^1 ({\widehat \gamma}^1 + {\widehat \gamma}^2) \right) \partial_{x^2} + ({\widehat \gamma}^1 + {\widehat \gamma}^2) \partial_{x^3} + \\
   &   & \frac{i}{\sqrt{k_{44}}} {\widehat \gamma}^3 \partial_{x^4} + \frac{c}{4\sqrt{k_{44}}} (i\stackrel{*}{\widehat \gamma^4} - \stackrel{*}{\widehat \gamma^3} - 4i\widehat \gamma^3)
\end{eqnarray*}
The spin symmetry operator (\protect\ref{sym2}) over the Killing-Yano tensor:
\begin{eqnarray*}
 L & = & \frac{i\exp(cx^4)}{2k_{13}} (\stackrel{*}{\widehat \gamma^1} - \stackrel{*}{\widehat \gamma^2}) \partial_{x^1} - \left(\frac{\sqrt{k_{44}}\exp(c x^4)}{ck_{13}} \stackrel{*}{\widehat \gamma^3} - ix^1 (\stackrel{*}{\widehat \gamma^1} + \stackrel{*}{\widehat \gamma^2}) \right) \partial_{x^2} - i(\stackrel{*}{\widehat \gamma^1} + \stackrel{*}{\widehat \gamma^2}) \partial_{x^3} + \\
   &   & \frac{i}{\sqrt{k_{44}}} \stackrel{*}{\widehat \gamma^4} \partial_{x^4} + \frac{c}{4\sqrt{k_{44}}} (i{\widehat \gamma}^3 + {\widehat \gamma}^4 - 4i\stackrel{*}{\widehat \gamma^4})
\end{eqnarray*}
A simple control gives $f^{in}f^j_{~n}=g^{ij}$ and we have the equation
$D^2=L^2$, so the operator $L$ is the second Dirac operator.
Solving the system (\protect\ref{equiv}) we find a matrix $S$ that gives
standard equivalence. The matrix has the following meaning:
\begin{displaymath}
S=\frac{1}{2}(1 - i\widehat\gamma - {\widehat\gamma}^{12} + i{\widehat\gamma}^{34}),
\end{displaymath}
it has the property $S^2=1$.
Operator $D^{(0)}_{-} = \{D,S\} = D + L = 2D^{(+)}$.

We shall investigate other metrics in the monograph \cite{Petrov} analogously.

Metric (33.32) in the monograph \cite[p 309]{Petrov}.
\begin{eqnarray}
ds^2 & = & k_{11} \exp(-2x^4) dx^{1^2} + k_{22} \exp(-4x^4) dx^{2^2} + 2k_{22} x^1 \exp(-4x^4) dx^2 dx^3 + \nonumber \\
     &   & (k_{22} x^{1^2} \exp(-4x^4) + k_{11} \exp(-2x^4)) dx^{3^2} + k_{44} dx^{4^2} \label{m33_32}
\end{eqnarray}
where $k_{ij}$ are arbitrary constants,
$\det(g_{ij})= k_{11}^2 k_{22} k_{44} \exp( - 8x^4)$,
and the space scalar curvature $R=-k_{22}/(2k_{11}^2) - 22/k_{44}$.

In the general case the Riemannian space (\protect\ref{m33_32}) does not
admit a Yano vector field and a Killing-Yano tensor field. However, if
$k_{22}k_{44} - 4 k_{11}^2 = 0$ there is a Killing-Yano tensor field
with non-zero components:
$f_{13} = a\exp(-2 x^4)$,
$f_{24} = 2a\exp(-2 x^4)$,
$f_{34} = 2a x^1 \exp(-2 x^4)$,
$a=ik_{11}$.

The Dirac operator in the space (\protect\ref{m33_32}):
\begin{eqnarray*}
 D & = & \frac{i\exp(x^4)}{\sqrt{k_{11}}} {\widehat \gamma}^2 \partial_{x^1} + \exp(x^4) \left(\frac{\sqrt{k_{44}}\exp(x^4)}{2k_{11}} {\widehat \gamma}^1 - \frac{ix^1}{\sqrt{k_{11}}} {\widehat \gamma}^3\right) \partial_{x^2} + \\
   &   & \frac{i\exp(x^4)}{\sqrt{k_{11}}} {\widehat \gamma}^3 \partial_{x^3} + \frac{i}{\sqrt{k_{44}}} {\widehat \gamma}^4 \partial_{x^4} - \frac{1}{2\sqrt{k_{44}}} (4i{\widehat \gamma}^4 - \stackrel{*}{\widehat \gamma^4})
\end{eqnarray*}
The second Dirac operator:
\begin{eqnarray*}
 L & = & \frac{i\exp(x^4)}{\sqrt{k_{11}}} \stackrel{*}{\widehat \gamma^3} \partial_{x^1} + i\exp(x^4)\left(\frac{\sqrt{k_{44}}\exp(x^4)}{2k_{11}} \stackrel{*}{\widehat \gamma^4} + \frac{x^1}{\sqrt{k_{11}}} \stackrel{*}{\widehat \gamma^2}\right) \partial_{x^2} - \\
   &   & \frac{i\exp(x^4)}{\sqrt{k_{11}}} \stackrel{*}{\widehat \gamma^2} \partial_{x^3} - \frac{1}{\sqrt{k_{44}}} \stackrel{*}{\widehat \gamma^1} \partial_{x^4} + \frac{1}{2\sqrt{k_{44}}} (4\stackrel{*}{\widehat \gamma^1} - i{\widehat \gamma}^1)
\end{eqnarray*}
The matrix $S=\frac{1}{2}(1 + i\widehat\gamma - {\widehat\gamma}^{14} - i{\widehat\gamma}^{23})$, $S^2 = 1$.
Operator $D^{(0)}_{-} = 2D^{(+)}$.

Metric (33.45) in the monograph \cite[p 312]{Petrov}.
\begin{eqnarray}
ds^2 & = & k_{11} \exp(x^4) dx^{1^2} + 2 k_{13} \exp(-x^4/2) dx^1 dx^3 + k_{22} \exp(-x^4) dx^{2^2} + \nonumber \\
     &   & 2 k_{22} x^1 \exp(-x^4) dx^2 dx^3 + (k_{22} x^{1^2} \exp(x^4) + k_{33}) \exp(-2x^4) dx^{3^2} + k_{44} dx^{4^2} \label{m33_45}
\end{eqnarray}
where $k_{ij}$ are constants, $\det (g_{ij}) = k_{22} k_{44} (k_{11}k_{33} - k_{13}^2) \exp(- 2x^4)$,
and the space scalar curvature
$R=(20 k_{11} k_{33} - 11k_{13}^2 + 4k_{22}k_{44})/(8k_{44}(k_{11}k_{33} - k_{13}^2))$.

The space (\protect\ref{m33_45}) admits five Killing vectors
$\xi^i_{(1)} = (0,1,0,0)$, $\xi^i_{(2)} = (0,0,1,0)$,
$\xi^i_{(3)} = (-1,x^3,0,0)$, $\xi^i_{(4)} = (-x^1/2,x^2/2,x^3,1)$,
$\xi^i_{(5)} = (k_1 \exp(x^4/2) + x^1 x^3 + x^2, x^2 x^3 - k_3 x^1 \exp(2x^4), x^{3^2} + k_3 \exp(2x^4), 2 x^3)$.

A constant $k_1,k_3$ lies under the conditions:
$k_{13} - k_1 k_{22} = 0$, $k_{11} + k_3 k_{22} = 0$, $k_1 k_{11} + 4k_3 k_{13} = 0$,
$k_1 k_{13} + 4k_3 k_{33} + 4k_{44} = 0$. Combining the first three equalities
above we find that $k_{11}k_{13}=0$. Both these constants cannot be zero
simultaneously otherwise the metric is degenarate, which is why one should
take two cases: a) $k_{11} = 0$, $k_{13} \ne 0$ and b) $k_{11} \ne 0$, $k_{13} = 0$.

The Riemannian space (\protect\ref{m33_45}) admits Killing-Yano tensor fields
in the following cases:

(a) $k_{11}=0$, $k_{33}=0$, $4k_{22}k_{44} + k_{13}^2=0$. Non-zero components are:
$f_{13} = 2a\exp(-x^4/2)$,
$f_{24} = a\exp(-x^4/2)$,
$f_{34} = a x^1 \exp(-x^4/2)$,
$a = ik_{13}/2$.
If a constant $k_{33} \ne 0$, our Riemannian space does not admit a
Killing-Yano tensor field.

(b) $k_{13} = 0$, $k_{11}k_{33} - k_{22}k_{44}=0$. Non-zero components are:
$f_{12} = ak_{11}$,
$f_{13} = ak_{11}x^1$,
$f_{34} = ak_{44}\exp(-x^4)$,
$a = \sqrt{k_{33}}/\sqrt{k_{44}} $.

{\it Case (a)} $k_{11}=0$, $k_{33}=0$, $4k_{22}k_{44} + k_{13}^2=0$.

The Dirac operator in the space (\protect\ref{m33_45}) has the form:
\begin{eqnarray*}
 D & = & ({\widehat \gamma}^1 + {\widehat \gamma}^2) \partial_{x^1} + \frac{\exp(x^4/2)}{k_{13}} \left(2 \sqrt{k_{44}} {\widehat \gamma}^3 - \frac{x^1}{2}({\widehat \gamma}^1 - {\widehat \gamma}^2) \right) \partial_{x^2} + \\
   &   & \frac{\exp(x^4/2)}{2k_{13}} \left({\widehat \gamma}^1 - {\widehat \gamma}^2\right) \partial_{x^3} + \frac{i}{\sqrt{k_{44}}} {\widehat \gamma}^4 \partial_{x^4} - \frac{1}{8\sqrt{k_{44}}} (4i{\widehat \gamma}^4 + i\stackrel{*}{\widehat \gamma^3} + \stackrel{*}{\widehat \gamma^4})
\end{eqnarray*}
The second Dirac operator:
\begin{eqnarray*}
 L & = & i(\stackrel{*}{\widehat \gamma^1} + \stackrel{*}{\widehat \gamma^2}) \partial_{x^1} - \frac{\exp(x^4/2)}{k_{13}} \left(2\sqrt{k_{44}} \stackrel{*}{\widehat \gamma^4} - \frac{ix^1}{2}(\stackrel{*}{\widehat \gamma^1} - \stackrel{*}{\widehat \gamma^2})\right) \partial_{x^2} - \\
   &   & \frac{i\exp(x^4/2)}{2k_{13}} (\stackrel{*}{\widehat \gamma^1} - \stackrel{*}{\widehat \gamma^2}) \partial_{x^3} + \frac{i}{\sqrt{k_{44}}} \stackrel{*}{\widehat \gamma^3} \partial_{x^4} + \frac{1}{8\sqrt{k_{44}}} ({\widehat \gamma^3} - i{\widehat \gamma}^4 - 4i\stackrel{*}{\widehat \gamma^3})
\end{eqnarray*}
Matrix $S=\frac{1}{2}(1-i\widehat\gamma+\widehat\gamma^{12}-i\widehat\gamma^{34})$, $S^2 = 1$.
Operator $D^{(0)}_{-} = 2D^{(+)}$.

{\it Case (b)} $k_{13} = 0$, $k_{11}k_{33} - k_{22}k_{44}=0$.

The Dirac operator has the form
\begin{eqnarray*}
 D & = & \frac{i\exp(-x^4/2)}{\sqrt{2k_{11}}} ({\widehat \gamma}^2 - {\widehat \gamma}^3) \partial_{x^1} + \frac{\exp(x^4)}{\sqrt{k_{33}}} \left(\frac{i\sqrt{k_{44}}\exp(-x^4/2)}{\sqrt{2k_{11}}} ({\widehat \gamma}^2 + {\widehat \gamma}^3) - x^1 {\widehat \gamma}^1 \right) \partial_{x^2} + \\
   &   & \frac{\exp(x^4)}{\sqrt{k_{33}}}{\widehat \gamma}^1 \partial_{x^3} + \frac{i}{\sqrt{k_{44}}} {\widehat \gamma}^4 \partial_{x^4} - \frac{1}{4\sqrt{k_{44}}}(2i{\widehat \gamma}^4 + \stackrel{*}{\widehat \gamma^4})
\end{eqnarray*}

The second Dirac operator:

\begin{eqnarray*}
 L & = & - \frac{i\exp(-x^4/2)}{\sqrt{2k_{11}}} (\stackrel{*}{\widehat \gamma^2} + \stackrel{*}{\widehat \gamma^3}) \partial_{x^1} + \frac{i\exp(x^4)}{\sqrt{k_{33}}} \left(\frac{\sqrt{k_{44}} \exp(-x^4/2)}{\sqrt{2k_{11}}} (\stackrel{*}{\widehat \gamma^2} - \stackrel{*}{\widehat \gamma^3}) + x^1 \stackrel{*}{\widehat \gamma^4} \right) \partial_{x^2} - \\
   &   & \frac{i\exp(x^4)}{\sqrt{k_{33}}} \stackrel{*}{\widehat \gamma^4} \partial_{x^3} + \frac{1}{\sqrt{k_{44}}} \stackrel{*}{\widehat \gamma^1} \partial_{x^4} - \frac{1}{4\sqrt{k_{44}}} (i{\widehat \gamma}^1 + 2 \stackrel{*}{\widehat \gamma^1})
\end{eqnarray*}

Matrix $S=\frac{1}{2}(1 + i\widehat\gamma + {\widehat\gamma}^{14} + i{\widehat\gamma}^{23})$, $S^2 = 1$.
Operator $D^{(0)}_{-} = 2D^{(+)}$.

Metric (33.46) in the monograph \cite[p 312]{Petrov}.
\begin{eqnarray}
ds^2 & = & k_{11} \exp(4x^4) dx^{1^2} + 2 k_{13} \exp(x^4) dx^1 dx^3 + k_{22} \exp(2x^4) dx^{2^2} + \nonumber \\
     &   & 2 k_{22} x^1 \exp(2x^4) dx^2 dx^3 + (k_{22} x^{1^2} \exp(4x^4) + k_{33}) \exp(-2x^4) dx^{3^2} + k_{44} dx^{4^2} \label{m33_46}
\end{eqnarray}
where $k_{ij}$ are constants, $\det (g_{ij})= k_{22} k_{44} (k_{11}k_{33} - k_{13}^2) \exp(4x^4)$,
and the space scalar curvature
$R = (20 k_{11} k_{33} - 11k_{13}^2 + k_{22}k_{44})/(2k_{44}(k_{11}k_{33} - k_{13}^2))$.

A space (\protect\ref{m33_46}) admits five Killing vectors, they have
the form $\xi^i_{(1)} = (0,1,0,0)$, $\xi^i_{(2)} = (0,0,1,0)$,
$\xi^i_{(3)} = (-1,x^3,0,0)$, $\xi^i_{(4)} = (-2x^1,-x^2,x^3,1)$,
$\xi^i_{(5)} = (k_1 \exp(-4x^4) + x^{1^2}, - k_3 x^1 \exp(-x^4), x^2 + k_3\exp(-x^4), - x^1)$.

The constants $k_1,k_3$ have the following conditions:
$k_{33} + k_1 k_{22} = 0$, $k_{13} - k_3 k_{22} = 0$, $4k_1 k_{13} + k_3 k_{33} = 0$,
$4k_1 k_{11} + k_3 k_{13} + k_{44} = 0$. Combining three equations,
we find that $k_{13}k_{33}=0$. Both the constants are non-zero simultaneously
so we shall have two cases:
(a) $k_{33} = 0$, $k_{13} \ne 0$;
(b) $k_{33} \ne 0$, $k_{13} = 0$.

The Riemannian space (\protect\ref{m33_46}) admits a Killing-Yano tensor
field in the following cases:

(a) $k_{33}=0$, $k_{11}=0$, $k_{22}k_{44} + k_{13}^2=0$. The non-zero components are:
$f_{13} = - a\exp(x^4)$,
$f_{24} = a\exp(x^4)$,
$f_{34} = ax^1 \exp(x^4)$,
$a=ik_{13}$.
If a constant $k_{11} \ne 0$, the Riemannian space (33.46) does not admit
a Killing-Yano tensor field.

(b) $k_{13} = 0$, $4k_{11}k_{33} - k_{22}k_{44}=0$. Non-zero components are:
$f_{14} = - a k_{44} \exp(2x^4)$,
$f_{23} = 2ak_{33}$,
$a=i\sqrt{k_{11}}/\sqrt{k_{44}}$.

{\it Case (a)} $k_{33}=0$, $k_{11}=0$, $k_{22}k_{44} + k_{13}^2=0$.

The Dirac operator in the space (\protect\ref{m33_46}):
\begin{eqnarray*}
 D & = & ({\widehat \gamma}^1 + {\widehat \gamma}^2) \partial_{x^1} + \frac{\exp(-x^4)}{k_{13}} \left( \sqrt{k_{44}} {\widehat \gamma}^3 - \frac{x^1}{2}({\widehat \gamma}^1 - {\widehat \gamma}^2) \right) \partial_{x^2} + \\
   &   & \frac{\exp(-x^4)}{2k_{13}} ({\widehat \gamma}^1 - {\widehat \gamma}^2)\partial_{x^3} + \frac{i}{\sqrt{k_{44}}} {\widehat \gamma}^4 \partial_{x^4} + \frac{1}{4\sqrt{k_{44}}} (4i{\widehat \gamma}^4 + i \stackrel{*}{\widehat \gamma^3} - \stackrel{*}{\widehat \gamma^4})
\end{eqnarray*}
The second Dirac operator:
\begin{eqnarray*}
 L & = & - i(\stackrel{*}{\widehat \gamma^1} + \stackrel{*}{\widehat \gamma^2}) \partial_{x^1} - \frac{\exp(-x^4)}{k_{13}} \left(\sqrt{k_{44}} \stackrel{*}{\widehat \gamma^4} + \frac{ix^1}{2}(\stackrel{*}{\widehat \gamma^1} - \stackrel{*}{\widehat \gamma^2}) \right) \partial_{x^2} + \\
   &   & \frac{i\exp(-x^4)}{2k_{13}} (\stackrel{*}{\widehat \gamma^1} - \stackrel{*}{\widehat \gamma^2}) \partial_{x^3} + \frac{i}{\sqrt{k_{44}}} \stackrel{*}{\widehat \gamma^3} \partial_{x^4} + \frac{1}{4\sqrt{k_{44}}} ({\widehat \gamma}^3 + i{\widehat \gamma}^4 + 4i\stackrel{*}{\widehat \gamma^3})
\end{eqnarray*}
The matrix $S=\frac{1}{2}(1 + i\widehat\gamma + {\widehat\gamma}^{12} + i{\widehat\gamma}^{34})$, $S^2 = 1$.
Operator $D^{(0)}_{-} = 2D^{(+)}$.

{\it Case (b)} $k_{13} = 0$, $4k_{11}k_{33} - k_{22}k_{44}=0$.

The Dirac operator in the space (\protect\ref{m33_46}):
\begin{eqnarray*}
 D & = & \frac{i\exp(-2x^4)}{\sqrt{2k_{11}}} ({\widehat \gamma}^2 - {\widehat \gamma}^3) \partial_{x^1} - \frac{\exp(-x^4)}{\sqrt{k_{33}}} \left(\exp(2 x^4) {\widehat \gamma}^1 x^1 - \frac{i\sqrt{k_{44}}}{2\sqrt{2k_{11}}} ({\widehat \gamma}^2 + {\widehat \gamma}^3) \right) \partial_{x^2} + \\
   &   & \frac{\exp(x^4)}{\sqrt{k_{33}}} {\widehat \gamma}^1 \partial_{x^3} + \frac{i}{\sqrt{k_{44}}} {\widehat \gamma}^4 \partial_{x^4} + \frac{1}{2\sqrt{k_{44}}} (2i{\widehat \gamma}^4 - \stackrel{*}{\widehat \gamma^4})
\end{eqnarray*}
The second Dirac operator:
\begin{eqnarray*}
 L & = & \frac{i\exp(-2x^4)}{\sqrt{k_{11}}} \stackrel{*}{\widehat \gamma^4} \partial_{x^1} - \frac{\exp(-x^4)}{\sqrt{k_{33}}} \left(\frac{i x^1\exp(2x^4)}{\sqrt{2}} (\stackrel{*}{\widehat \gamma^2} + \stackrel{*}{\widehat \gamma^3}) + \frac{\sqrt{k_{44}}}{2\sqrt{k_{11}}} \stackrel{*}{\widehat \gamma^1}\right) \partial_{x^2} + \\
   &   & \frac{i\exp(x^4)}{\sqrt{2 k_{33}}} (\stackrel{*}{\widehat \gamma^2} + \stackrel{*}{\widehat \gamma^3}) \partial_{x^3} - \frac{i}{\sqrt{2k_{44}}} (\stackrel{*}{\widehat \gamma^2} - \stackrel{*}{\widehat \gamma^3}) \partial_{x^4} - \frac{1}{2\sqrt{2k_{44}}} ({\widehat \gamma}^2 - {\widehat \gamma}^3 + 2i \stackrel{*}{\widehat \gamma^2} - 2i \stackrel{*}{\widehat \gamma^3})
\end{eqnarray*}
Matrix $S=\frac{i}{2\sqrt{2}}(i\sqrt{2} + \sqrt{2}\widehat\gamma + i{\widehat\gamma}^{12} + i{\widehat\gamma}^{13} - {\widehat\gamma}^{24} + {\widehat\gamma}^{34})$, $S^2 = 1$.
Operator $D^{(0)}_{-} = - 2D^{(+)}$.

Let us give an example of gravitational fields with equation
(\protect\ref{form}) under the condition $\lambda = 0$. In Petrov's
classification there are only two examples of Riemannian spaces with this
property.

Metric (33.49) in the monograph \cite[p 313]{Petrov}.
\begin{eqnarray}
ds^2 & = & 2k_{12} dx^1 dx^2 - ((k_{12} - k_{33})x^{4^2} + 2k_{23}x^4 - k_{22}) dx^{2^2} + \nonumber \\
     &   & 2 ((k_{12} - k_{33}) x^4 + k_{23}) dx^2 dx^3 + k_{33} dx^{3^2} + k_{44} dx^{4^2} \label{m33_49}
\end{eqnarray}
where $k_{ij}$ are constants, $\det (g_{ij}) = - k_{12}^2 k_{33} k_{44}$,
space scalar curvature is $R = 0$. The space metric (\protect\ref{m33_49})
satisfies Einstein's vacuum equations (\protect\ref{einst}) ($\Lambda = 0$)
under the condition $k_{12} = \pm k_{33}$.
The space (\protect\ref{m33_49}) admits two Killing-Yano tensors
$f_{23} = i(\alpha\exp(i\omega x^2) - \beta\exp(-i\omega x^2))\sqrt{k_{33}}/\sqrt{k_{44}}$,
$f_{24} = \alpha\exp(i\omega x^2) + \beta\exp(-i\omega x^2)$. Here
$\alpha,\beta$ are constants, $\omega = (k_{12} - k_{33})/(2\sqrt{k_{33}k_{44}})$.

Metric (33.50) in the monograph \cite[p 313]{Petrov}.
\begin{eqnarray}
ds^2 & = & k_{11} dx^{1^2} - 2k_{13}x^4 dx^1dx^2 + 2 k_{13} dx^1 dx^3 + k_{22} dx^{2^2} + k_{44} dx^{4^2} \label{m33_50}
\end{eqnarray}
where $k_{ij}$ are constants, $\det (g_{ij}) = - k_{13}^2 k_{22} k_{44}$
and the space scalar curvature $R = 0$. The space metric (\protect\ref{m33_50})
does not satisfy Einstein's equations (\protect\ref{einst}) for any constants.
The space (\protect\ref{m33_50}) admits two Killing-Yano tensors
$f_{12} = \alpha\exp(i\omega x^2) + \beta\exp(-i\omega x^2)$,
$f_{14} = 2i\omega(\alpha\exp(i\omega x^2) - \beta\exp(-i\omega x^2)) k_{44}/k_{13}$,
$\alpha,\beta$ are constants, $\omega^2 = k_{13}^2/(4 k_{22}k_{44})$.

The Killing-Yano tensor fields for metrics (\protect\ref{m33_49}) and
(\protect\ref{m33_50}) have the following property: $f^{in}f^{j}_{~n} = 0$ and
the symmetry operators (\protect\ref{sym2}) constructed from the tensor
fields satisfy the condition $L^2=0$ correspondingly.

\section{Solution of the Einstein equation and Petrov types \label{s4}}

There are only four metrics (\protect\ref{m33_31})-(\protect\ref{m33_46})
with five dimensional motion group where the second Dirac operator arises.
Metrics (\protect\ref{m33_31}), (\protect\ref{m33_32}),
(\protect\ref{m33_45}), case ({\it a}) and (\protect\ref{m33_46}), case ({\it b})
satisfy Einstein's equations with the $\Lambda$-member
\begin{equation}
R_{ij} - \frac{R}{2} g_{ij} = \Lambda g_{ij} \label{einst}
\end{equation}
under the following conditions:
 $\Lambda = 3c^2/(2k_{44})$,
 $\Lambda = 6/k_{44}$,
 $\Lambda = 3/(8k_{44})$,
 $\Lambda = 3/(2k_{44})$
correspondingly.
Spaces (\protect\ref{m33_31}), (\protect\ref{m33_32}), (\protect\ref{m33_45}),
case ({\it a}), (\protect\ref{m33_46}), case ({\it a}) belong to the Petrov
type D, and the rest metrics to type I by Petrov's classification. From the
Newman-Penrose equations we find independent components of the Weyl tensor
defined by five complex scalars $\psi_0$, $\psi_1$, $\psi_2$, $\psi_3$, $\psi_4$.
The Petrov type is then established from analysis of the roots of the
equation $\psi_4x^4+4\psi_3x^3+6\psi_2x^2+4\psi_1x+\psi_0=0$ \cite[p 74]{Chandr}.

I will now give some remarks concerning the discussed metrics.
Metrics (\protect\ref{m33_45}) and (\protect\ref{m33_46}) belong to
the non-St\"{a}ckel type, i.e. it is impossible to find the solution of the
Klein-Fock and Dirac equations according to the method of separation
of variables in these spaces.
Metrics (\protect\ref{m33_31}) and (\protect\ref{m33_32}) belong to
the St\"{a}ckel type, but it is impossible to find the solution of the Dirac
equation with the method of separation of variables in this spaces too.
However, in all discussed metrcis the Dirac equation is integrated so we can
find the exact solution using some other method, the results are given in \cite{Shir2}-\cite{Var2}.
The identities $D^2 = L^2$ were accidentally found when attempts to integrate
the Dirac equation were made with the help of the method of separation of
variables and the additional symmetry operators (\protect\ref{sym2}) and (\protect\ref{sym3}).

\section{Hypothesis concerning the number of second Dirac operators \label{s5}}

Besides all discussed metrics the author knows some more examples of the
Riemannian spaces (even Ricci flat), where a second Dirac operator
arises. In the author's opinion there is an interesting feature that
the number of second Dirac operators is always odd, i.e. the common number
of Dirac operators is always even (two or four). However, the author does not
have a proof of this working hypothesis, but does not have any examples
contradicting it.
It is necessary to note that in all examples the number of operators
(\protect\ref{sym2}) under the condition $L^2=0$ (the case $\lambda=0$ in
equation (\protect\ref{form})) is even as well. This hypothesis also needs
to be explained. There are some examples where both the second Dirac operator
and two operators (\protect\ref{sym2}) under the condition $L^2=0$ arise.
The roots of these phenomena seem to lie in supersymmetry theory.

\section{Conclusion}

The condition (\protect\ref{kiling3'}) in the general case defines a
Riemannian space class where the identities $D^2=L^2$ exist between the Dirac
operator $D$ and the symmetry operator $L$ (\protect\ref{sym2}). In this case
we interpret the operator $L$ as a second Dirac operator. It is a main result
of this paper. In our opinion the existence of the second Dirac operator is
a property of the gravitational field itself. The state of the spin particle
in this gravitational field is degenarate. There are two solutions
$\Psi^{(\pm)}$ of the initial Dirac equation (\protect\ref{dirac}).
An equivalence $SL=DS$ between the operators $D$ and $L$ is
important and always works if we choose a spin connection in the form
(\protect\ref{sviaz}). We give a physical interpretation of the value $S$
as a parity analogue. In section \ref{s3} we give some examples of
gravitational fields where the second Dirac operator exists.

In conclusion, I would like to note that in flat space and de Sitter space
all the Killing-Yano fields are well known. The control of conditions
(\protect\ref{kiling3'}) in these spaces is negative.
So in flat space and de Sitter space a second Dirac operator does not exist.
This fact is well known in the literature.

One can use the results of this paper to find the exact solutions of the
Dirac equation in Riemannian space of type (\protect\ref{kiling3'}).

\section*{Acknowledgements}

I am very grateful to Professors A K Guts, I V Shirokov and I L Buchbinder
for their useful remarks. I would like to thank A V Vikulova for help in the
translation of this paper.

\newpage
\section*{Appendix A. Dirac matrices}

Some equivalent representations are known for Dirac matrices
\cite{Bjorken,Itzykson}. We use a standard representation,

\begin{equation}
{\widehat \gamma}^1 = \left (
\begin{array}{cc}
{\widehat E}_2 & 0 \\
 0 & - {\widehat E}_2
\end{array} \right),
{\widehat \gamma}^2 = \left (
\begin{array}{cc}
 0 & \sigma_x \\
 -\sigma_x & 0
\end{array} \right),
\end{equation}
\begin{equation}
{\widehat \gamma}^3 = \left (
\begin{array}{cc}
 0 & \sigma_y \\
 -\sigma_y & 0
\end{array} \right),
{\widehat \gamma}^4 = \left (
\begin{array}{cc}
 0 & \sigma_z \\
 -\sigma_z & 0
\end{array} \right).
\end{equation}

Here ${\widehat E}_2$ is a unit $2 \times 2$ matrix and
$\sigma_x$, $\sigma_y$, $\sigma_y$ are Pauli matrices.

\begin{equation}
{\widehat E}_2 = \left (
\begin{array}{rr}
 1 & 0 \\
 0 & 1
\end{array} \right),
\sigma_x = \left (
\begin{array}{rr}
 0 & 1 \\
 1 & 0
\end{array} \right),
\end{equation}
\begin{equation}
\sigma_y = \left (
\begin{array}{rr}
 0 & -i \\
 i & 0
\end{array} \right),
\sigma_z = \left (
\begin{array}{rr}
 1 & 0 \\
 0 & -1
\end{array} \right).
\end{equation}

One can construct matrices $\stackrel{*}{\widehat \gamma^i}$,
${\widehat \gamma^{ij}}$ and ${\widehat \gamma}$ with the help of the
following equations:
$\stackrel{*}{\widehat \gamma^1} = - {\widehat\gamma^2}{\widehat\gamma^3}{\widehat\gamma^4}$,
$\stackrel{*}{\widehat \gamma^2} = - {\widehat\gamma^1}{\widehat\gamma^3}{\widehat\gamma^4}$,
$\stackrel{*}{\widehat \gamma^3} =   {\widehat\gamma^1}{\widehat\gamma^2}{\widehat\gamma^4}$,
$\stackrel{*}{\widehat \gamma^4} = - {\widehat\gamma^1}{\widehat\gamma^2}{\widehat\gamma^3}$,
${\widehat \gamma^{ij}} = \frac{1}{2} [\widehat \gamma^i,\widehat \gamma^j]$,
${\widehat \gamma} = - \widehat \gamma^1\widehat \gamma^2\widehat\gamma^3\widehat\gamma^4$.

\section*{Appendix B. Tetrads}

The tetrad for the metric (\protect\ref{m33_31}) is

$e_{(1)}^i = \left(\exp(c x^4), - 2k_{13} x^1,2k_{13},0 \right)/(2k_{13})$;

$e_{(2)}^i = \left( - \exp(c x^4), - 2k_{13}x^1,2k_{13},0 \right)/(2k_{13})$;

$e_{(3)}^i = \left(0,0,0,i\right)/\sqrt{k_{44}}$;

$e_{(4)}^i = \left(0,i\exp(c x^4),0,0 \right)/\sqrt{k_{22}}$.

\vspace{5mm}
The tetrad for the metric (\protect\ref{m33_32}) is

$e_{(1)}^i = \left(0,\exp(2 x^4),0,0 \right)/\sqrt{k_{22}}$;

$e_{(2)}^i = \left(i\exp(x^4),0,0,0 \right)/\sqrt{k_{11}}$;

$e_{(3)}^i = \left(0, - ix^1\exp(x^4),i\exp(x^4),0 \right)/\sqrt{k_{11}}$;

$e_{(4)}^i = \left(0,0,0,i\right)/\sqrt{k_{44}}$.

\vspace{5mm}
The tetrad for the metric (\protect\ref{m33_45}) is

{\it Case (a)}

$e^i_{(1)} = \left(2k_{13},-x^1\exp(x^4/2),\exp(x^4/2),0 \right)/(2k_{13})$;

$e^i_{(2)} = \left(2k_{13},x^1\exp(x^4/2),-\exp(x^4/2),0 \right)/(2k_{13})$;

$e^i_{(3)} = \left(0,2\exp(x^4/2)\sqrt{k_{44}},0,0 \right)/k_{13}$;

$e^i_{(4)} = \left(0,0,0,i\right)/\sqrt{k_{44}}$.

\vspace{5mm}
{\it Case (b)}

$e^i_{(1)} = \left(0,-x^1\exp(x^4),\exp(x^4),0 \right)/\sqrt{k_{33}}$;

$e^i_{(2)} = \left(i\exp(-x^4/2),i\exp(x^4/2)\sqrt{k_{44}}/\sqrt{k_{33}},0,0 \right)/\sqrt{2k_{11}}$;

$e^i_{(3)} = \left(-i\exp(-x^4/2),i\exp(x^4/2)\sqrt{k_{44}}/\sqrt{k_{33}},0,0 \right)/\sqrt{2k_{11}}$;

$e^i_{(4)} = \left(0,0,0,i\right)/\sqrt{k_{44}}$.

\vspace{5mm}
The tetrad for the metric (\protect\ref{m33_46}) is

{\it Case (a)}

$e^i_{(1)} = \left(2k_{13},-x^1\exp(-x^4),\exp(-x^4),0 \right)/(2k_{13})$;

$e^i_{(2)} = \left(2k_{13},x^1\exp(-x^4),-\exp(-x^4),0 \right)/(2k_{13})$;

$e^i_{(3)} = \left(0,\exp(-x^4)\sqrt{k_{44}},0,0 \right)/k_{13}$;

$e^i_{(4)} = \left(0,0,0,i\right)/\sqrt{k_{44}}$.

\vspace{5mm}
{\it Case (b)}

$e_{(1)}^i = \left( 0, - \exp(x^4) x^1, \exp(x^4),0 \right)/\sqrt{k_{33}}$;

$e_{(2)}^i = \left( i\exp(-2 x^4),i\sqrt{k_{44}}\exp(-x^4)/(2 \sqrt{k_{33}}),0,0 \right)/\sqrt{2k_{11}}$;

$e_{(3)}^i = \left( - i\exp(-2 x^4),i\sqrt{k_{44}}\exp(-x^4)/(2 \sqrt{k_{33}}),0,0 \right)/\sqrt{2k_{11}}$;

$e_{(4)}^i = \left( 0,0,0,i\right)/\sqrt{k_{44}}$.

\vspace{1cm}


\end{document}